\documentclass[journal=jacsat,manuscript=article]{achemso}

\usepackage[version=3]{mhchem} 
\usepackage{xcolor}
\usepackage{adjustbox}
\usepackage{graphics}

\title{Inverse Design of Next-generation Superconductors Using Data-driven Deep Generative Models }% 

\author{Daniel Wines}%
 \email{daniel.wines@nist.gov}
\affiliation{%
 Material Measurement Laboratory, National Institute of Standards and Technology,
Gaithersburg, Maryland 20899, USA 
}%

\author{Tian Xie}%

\affiliation{%
 Microsoft Research AI4Science, Cambridge, United Kingdom CB1 2FB}%

\author{Kamal Choudhary}
 \affiliation{%
 Material Measurement Laboratory, National Institute of Standards and Technology,
Gaithersburg, Maryland 20899, USA 
}%
 \alsoaffiliation{%
DeepMaterials LLC, Silver Spring, MD 20906, USA}

\begin{document}

\begin{abstract}

Finding new superconductors with a high critical temperature ($T_c$) has been a challenging task due to computational and experimental costs. We present a diffusion model inspired by the computer vision community to generate new superconductors with unique structures and chemical compositions. Specifically, we used a crystal diffusion variational autoencoder (CDVAE) along with atomistic line graph neural network (ALIGNN) pretrained models and the Joint Automated Repository for Various Integrated Simulations (JARVIS) superconducting database of density functional theory (DFT) calculations to generate new superconductors with a high success rate. We started with a DFT dataset of $\approx$1000 superconducting materials to train the diffusion model. We used the model to generate 3000 new structures, which along with pre-trained ALIGNN screening results in 61 candidates. For the top candidates, we performed DFT calculations for validation. Such approaches go beyond the funnel-like materials design approaches and allow for the inverse design of next-generation materials.

\end{abstract}


%keywords{Suggested keywords}%Use showkeys class option if keyword
                              %display desired
\maketitle

After superconductivity was discovered in 1911 by Onnes \cite{kamerlingh1911resistance}, the efforts to identify novel superconducting materials with high transition temperatures ($T_c$) has been an intense area of research in materials science and condensed matter physics \cite{poole2013superconductivity,rogalla2011100}. There have been systematic computational efforts to identify Bardeen–Cooper–Schrieffer (BCS) conventional superconductors \cite{cooper2010bcs, giustino2017electron} with high-$T_c$ prior to costly experimental investigation \cite{bulksc,2dsc,PhysRevB.104.054501}, where density functional theory-perturbation theory (DFT-PT) calculations have been performed to obtain the electron-phonon coupling (EPC) parameters. In addition, various machine learning approaches have been utilized to accelerate the search for high-$T_c$ superconductors \cite{bulksc,ml-sc1,ml-sc2,ml-sc3,C8ME00012C,ROTER20201353689,ml-sc4}. However, these typical funnel-like funnel screening-based approaches are not sufficient for inverse materials design, where instead of engineering from structure to property, the goal is to engineer from a target property to crystal structure.

The inverse design of novel materials becomes a difficult problem to solve due to the fact that there are an infinite number of possible materials with varying properties that are dependent on chemical composition and crystal structure. In order for inverse design to be successful, an algorithm that can successfully create new candidate structures based on a high quality and diverse dataset of materials data is required. There currently exists abundant sources of high quality calculated material properties in databases such as Materials Project \cite{doi:10.1063/1.4812323}, Open Quantum Materials Database (OQMD) \cite{oqmd,oqmd-2}, C2DB \cite{Haastrup_2018,Gjerding_2021} and our own JARVIS \cite{choudhary2020joint} database, which all contain millions of high-throughput \cite{ONG2013314,choudhary2020joint} DFT calculations. The main obstacle in inverse design has been the method or algorithm used to generate new candidate materials. To circumvent this obstacle, generative machine learning algorithms can be used to successfully design new candidate structures. With the popularity of generative tools such as DALL-E \cite{ramesh2021zeroshot}, which uses deep learning models to generate digital images from natural language descriptions (prompts), the interest in using deep generative models for scientific applications has immensely increased \cite{wu2022protein,https://doi.org/10.48550/arxiv.2110.06197,cdvae-2d,PhysRevMaterials.7.014007}. One of the main challenges with generative models for periodic materials stems from the creation of representations that are translationally and rotationally invariant \cite{Fung_2022,D0SC00594K,genmod,genmod2,genmod3,NOH20191370,https://doi.org/10.1002/advs.202100566,REN2022314}.

In a recent work, Xie et al. developed a crystal diffusion variational autoencoder (CDVAE) model\cite{https://doi.org/10.48550/arxiv.2110.06197} (\url{https://github.com/txie-93/cdvae}) for periodic structure generation. The CDVAE consists of a variational autoencoder \cite{https://doi.org/10.48550/arxiv.1312.6114} and a diffusion model \cite{pmlr-v37-sohl-dickstein15,https://doi.org/10.48550/arxiv.1907.05600,sohldickstein2015deep} that works directly with the atomic coordinates of the structures and uses an equivariant graph neural network \cite{eq-gnn} to ensure invariance without the need for representations such as graphs or descriptors. The CDVAE consists of three simultaneously trained networks: 1) the encoder, which encodes onto the latent space, 2) the property predictor, which samples the latent space and predicts a structure and composition, and 3) the decoder, which is a diffusion model that denoises the randomly initialized atom types into a material that is similar to the training set distribution. More details of the CDVAE method can be found in Xie et al. \cite{https://doi.org/10.48550/arxiv.2110.06197}. A recent work by Lyngby and Thygesen \cite{cdvae-2d} successfully applied the CDVAE model to discover new, stable 2D materials and vastly expand the space of 2D materials (on the order of thousands). Another recent work by Moustafa et al. \cite{PhysRevMaterials.7.014007} used the CDVAE model to discover more than 500 new stable one-dimensional materials. In this work, we trained a CDVAE model (optimizing $T_c$ in the latent space) with DFT computed data of 1058 superconducting materials from the JARVIS database and generated thousands of new candidate superconductors. We screened these
candidate structures further by predicting the properties using pretrained deep learning models for fast predictions. After narrowing down the pool of potential candidate superconductors, we performed DFT calculations to verify our predictions and assessed the dynamical and thermodynamic stability of the newly predicted materials.

We utilized the atomistic line graph neural network (ALIGNN) \cite{choudhary2021atomistic} (\url{https://github.com/usnistgov/alignn}) to make deep learning predictions of the superconducting properties of each material. This pretrained model for superconducting properties was specifically developed in ref. \cite{bulksc}. ALIGNN is implemented in the deep graph library \cite{wang2019deep} and PyTorch \cite{paszke2019pytorch}. In the ALIGNN framework, a structure is represented as a graph, where the elements are nodes and bonds are edges. Nine input features are assigned to each node in the graph. These features include first ionization energy, electron affinity, electronegativity, block, valence electrons, group number, covalent radius, and atomic volume. The edge features are the bond distances, where the cutoff for the radial basis function is 8 $\textrm{\AA}$. A 12-nearest neighbor periodic graph construction is used. The line graph is constructed from the atomistic graph using bond distances as nodes and bond angles as edge features. Edge-gated graph convolution is used for updating the nodes and edge features by use of a propagation function. An edge-gated graph convolution on the bond graph with an edge-gated convolution on the line graph is what composes one layer. Bond messages are produced from the line graph convolution that propogate to the atomistic graph, where the bond features and atom features are further updated. With regards to predicting superconducting properties, we used a batch size of 16, 90:5:5 split and training for 300 epochs, where the test set was not used at all during training. We kept the hyperparameters of the model the same as the original ALIGNN paper \cite{choudhary2021atomistic}.

We utilize the publicly available JARVIS \cite{choudhary2020joint} infrastructure for our DFT and deep learning goals mentioned above. JARVIS (Joint Automated Repository for Various Integrated Simulations, \url{https://jarvis.nist.gov/}) is a collection of databases and tools to automate materials design using classical force-field, density functional theory, machine learning calculations and experiments. JARVIS-DFT is a density functional theory based database of over 75000 materials with several material properties such as formation energy, band gap with different level of theories \cite{choudhary2018computational}, solar-cell efficiency \cite{choudhary2019accelerated}, topological spin-orbit coupling spillage \cite{choudhary2021high,choudhary2019high,choudhary2020computational}, elastic tensors \cite{choudhary2018elastic}, dielectric tensors, piezoelectric tensors, infrared and Raman spectrum \cite{choudhary2020high}, electric field gradients \cite{choudhary2020density}, exfoliation energies \cite{choudhary2017high}, two-dimensional (2D) magnets \cite{crx3-qmc}, and bulk \cite{bulksc} and 2D superconductors \cite{2dsc}, all with stringent DFT-convergence setup \cite{choudhary2019convergence}. Our ALIGNN and CDVAE models were trained on the 1058 DFT calculations of superconducting properties presented in ref. \cite{bulksc}.

To verify top candidate superconductors, we followed the workflow used to generate the training data (in ref. \cite{bulksc}), where we performed EPC calculations using non-spin polarized DFT-PT \cite{baroni1987green,gonze1995perturbation} (using the interpolated/Gaussian broadening method \cite{wierzbowska2005origins}) with the Quantum Espresso (QE) software package \cite{giannozzi2009quantum}, PBEsol functional \cite{perdew2008restoring}, and the GBRV \cite{garrity2014pseudopotentials} pseudopotentials. The EPC parameter is derived from spectral function ${\alpha}^2 F(\omega)$ which is calculated as follows:

\begin{equation} 
{\alpha}^2 F(\omega)=\frac{1}{2{\pi}N({\epsilon_F})}\sum_{qj}\frac{\gamma_{qj}}{\omega_{qj}}\delta(\omega-\omega_{qj})w(q)
\end{equation} 
where $\omega_{qj}$ is the mode frequency, $N({\epsilon_F})$ is the DOS at the Fermi level ${\epsilon_F}$, $\delta$ is the Dirac-delta function, $w(q)$ is the weight of the $q$ point,  $\gamma_{qj}$ is the linewidth of a phonon mode $j$ at wave vector $q$ and is given by:

\begin{equation} 
\gamma_{qj}=2\pi \omega_{qj} \sum_{nm} \int \frac{d^3k}{\Omega_{BZ}}|g_{kn,k+qm}^j|^2 \delta (\epsilon_{kn}-\epsilon_F) \delta(\epsilon_{k+qm}-\epsilon_F)
\end{equation} 
Here, the integral is over the first Brillouin zone, $\epsilon_{kn}$  and $\epsilon_{k+qm}$ are the DFT eigenvalues with wavevector $k$ and $k+q$ within the $n$th and $m$th bands respectively, $g_{kn,k+qm}^j$ is the electron-phonon matrix element. $\gamma_{qj}$ is related to the mode EPC parameter $\lambda_{qj}$ by:

\begin{equation} 
\lambda_{qj}=\frac {\gamma_{qj}}{\pi hN(\epsilon_F)\omega_{qj}^2}
\end{equation} 
Now, the EPC parameter is given by:

\begin{equation} 
\lambda=2\int \frac{\alpha^2F(\omega)}{\omega}d\omega=\sum_{qj}\lambda_{qj}w(q)
\end{equation} 
with $w(q)$ as the weight of a $q$ point. The superconducting transition temperature, $T_c$ can then be approximated using McMillan-Allen-Dynes \cite{mcmillan1968transition} equation as follows:

\begin{equation}
T_c=\frac{\omega_{log}}{1.2}\exp\bigg[-\frac{1.04(1+\lambda)}{\lambda-\mu^*(1+0.62\lambda)}\bigg]\label{eq:mad}
\end{equation}
where
\begin{equation} 
\omega_{log}=\exp\bigg[\frac{\int d\omega \frac{\alpha^2F(\omega)}{\omega}\ln\omega}{\int d\omega \frac{\alpha^2F(\omega)}{\omega}}\bigg]
\end{equation} 
In Eq.~\ref{eq:mad}, the parameter $\mu^*$ is the effective Coulomb potential parameter, which we take as 0.1. It is important to note that the robustness of this workflow was heavily benchmarked against experimental data and higher levels of theory in ref. \cite{bulksc} and \cite{2dsc}, which indicates that the training data for these deep learning models is of high quality, given the level of theory used to produce the data.  

\begin{figure}[h]
    \centering
    \includegraphics[trim={0. 0cm 0 0cm},clip,width=0.55\textwidth]{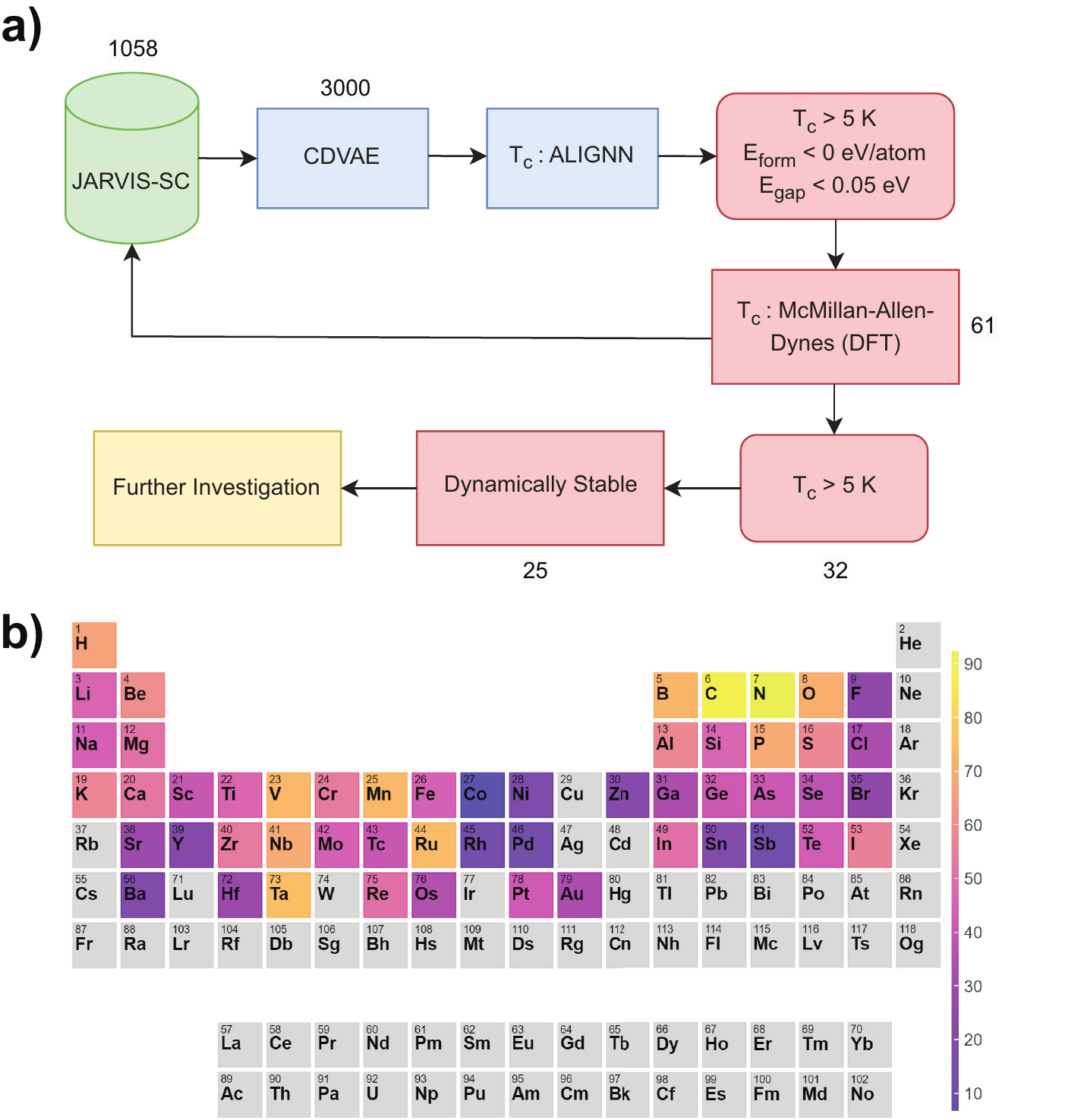}
    \caption{a) The full inverse design workflow for new superconductors using DFT, ALIGNN and the CDVAE generative model and b) the probability that compounds containing a given element in the CDVAE generated structures have an ALIGNN predicted $T_c$ $>$ 3 K.  }
    \label{cdvaeworkflow}
\end{figure}

The full inverse design workflow proposed in this work is depicted in Fig. \ref{cdvaeworkflow}a). The first step of this workflow involved training the CDVAE model on the 1058 DFT calculations in the JARVIS-SC (superconducting) database (from ref. \cite{bulksc}). With regards to the CDVAE model for inverse design, the target property to optimize was $T_c$ for new candidate superconductors. After training on the JARVIS-SC data, we generated 3000 candidate materials using the CDVAE model, where $T_c$ was optimized in the latent space. To ensure that the CDVAE model optimized $T_c$ in the latent space and actually ``learned" superconductivity from the JARVIS-SC training data, we analyzed the property loss function (which is depicted in Fig. S1). As seen in Fig. S1, the loss function is sharply decreasing and converging around 1.2 K. This converged value is comparable to the mean absolute error (MAE) achieved in the ALIGNN model for $T_c$ reported in ref. \cite{bulksc}.

With such a large amount of crystal structures to investigate, it is impracticable to perform DFT calculations of the EPC workflow for all of the candidates. For this reason, we used our deep learning property prediction tool (ALIGNN) to screen all 3000 candidate materials. ALIGNN has demonstrated success for predicting properties such as formation energy and band gap and more recently superconducting $T_c$. The screening criteria we established for further investigation of these candidates was an ALIGNN predicted: $T_c$ $>$ 5 K, $E_\textrm{form}$ $<$ 0 eV/atom, and $E_\textrm{gap}$ $<$ 0.05 eV. This criterion is based on the fact that we want to further investigate superconducting materials which have a high $T_c$, are potentially stable, and are metallic (high density of states at the Fermi level). Although negative formation energy is a stringent requirement for stability, investigations of dynamical and thermodynamic stability are still needed to confirm stability (phonon spectrum and energy above the convex hull). After performing this screening with ALIGNN, we found 61 materials which fit this criterion. We then went on to perform the full DFT-EPC workflow for these 61 materials and computed the $T_c$ using the McMillan-Allen-Dynes equation. After computing the $T_c$ of these materials with DFT, we found that 32 structures have a $T_c$ above 5 K. Upon investigation of the phonon density of states and chemical composition, we find that 7 of these structures have negative phonon frequencies, indicating dynamical instability. The remaining 25 candidate superconductors are presented in Table \ref{topstructures}.

\begin{figure}[h]
    \centering
    \includegraphics[trim={0. 0cm 0 0cm},clip,width=0.65\textwidth]{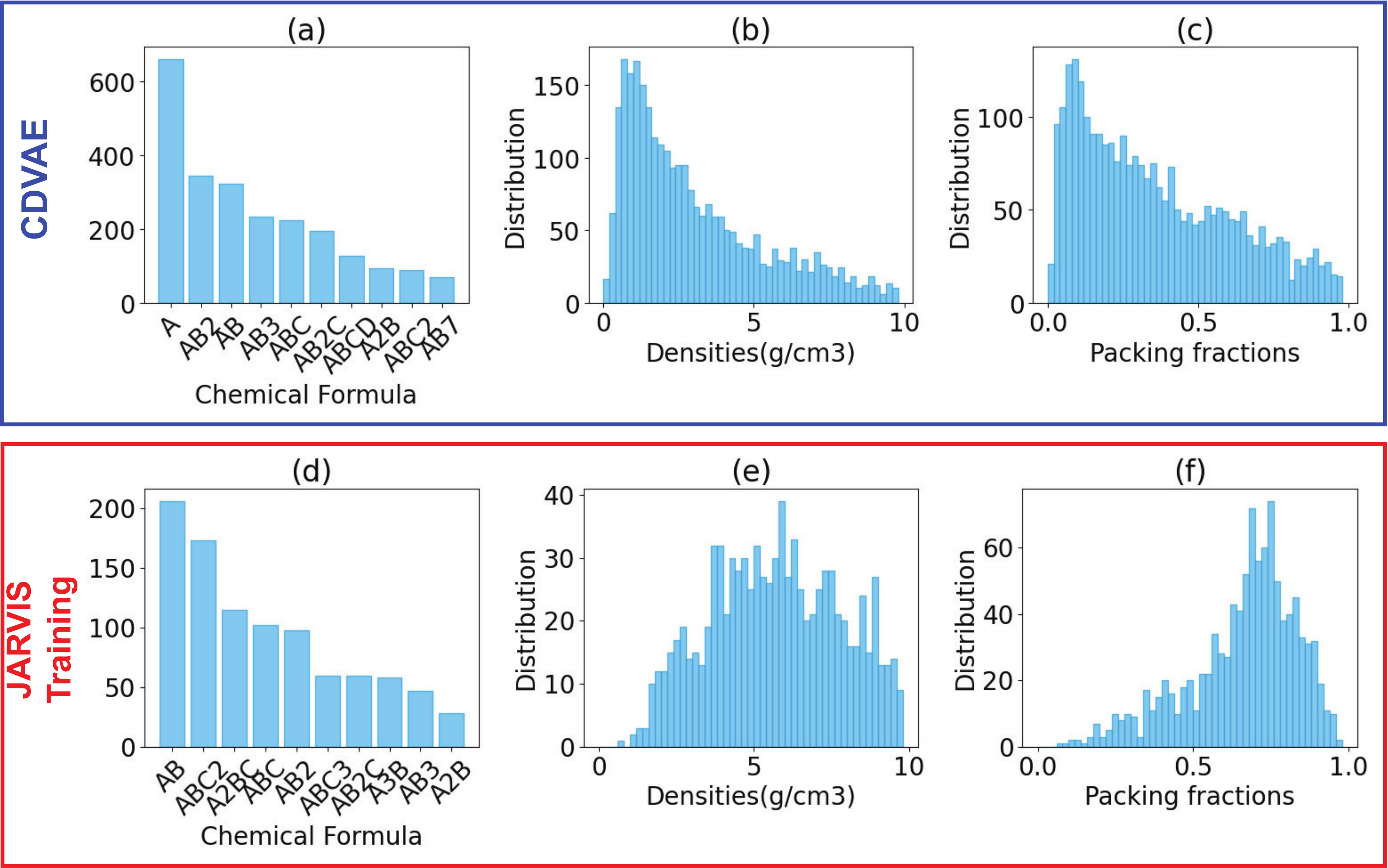}
    \caption{The chemical formula, distribution of densities, and distribution of packing fractions for a) - c) the 3000 CDVAE generated structures and for d) - f) the 1058 structures used for training from the JARVIS-DFT database.}
    \label{cdvaedist}
\end{figure}

Before discussing the more detailed results of the 25 new candidate superconducting materials generated by our workflow depicted in Fig. \ref{cdvaeworkflow}a), we will discuss the intermediate steps of the workflow in more detail. The CDVAE model has the capability to produce new structures with chemical and structural diversity. Fig. \ref{cdvaeworkflow}b) depicts the probability that compounds containing a given element in the 3000 CDVAE generated structures have an ALIGNN predicted $T_c$ above 3 K (displayed as a heat map overlayed on the periodic table). We observe that compounds containing C, N, B, O, V, Mn, Nb, Ru, and Ta are the most abundant with regards to structures that have a higher predicted $T_c$ with ALIGNN. This is not surprising, due to the fact that several of the materials in the DFT training set that have a high value of $T_c$ contain such elements (see ref. \cite{bulksc} for details). 

Although the abundance of elements in the CDVAE structures (with an ALIGNN predicted $T_c$ $>$ 3 K) follows similar trends to the training data, the CDVAE candidate structures are not restricted to stoichiometries and crystal structures already present in the training data. This is illustrated by Fig. \ref{cdvaedist}, where the chemical formula, distribution of densities, and distribution of packing fractions for a) - c) the 3000 CDVAE generated structures and for d) - f) the 1058 structures used for training from the JARVIS-DFT database is depicted. A comparison of Fig. \ref{cdvaedist}a) and \ref{cdvaedist}d) emphasizes the difference in the stoichiometry between the CDVAE generated structures and the training data from JARVIS. For example, in the JARVIS training set, a majority of the data contains only three different atomic species (A, B, and C) and the unit cells have 5 formula units (see Fig. \ref{cdvaedist}d)). In contrast, the CDVAE generated data based on the JARVIS training set see Fig. \ref{cdvaedist}a)) has a number of structures that contain four atomic species (A, B, C, D) and unit cells that have a larger number of formula units (i.e., AB$_7$). In addition, the CDVAE generated data contains a substantial number of monoelemental structures, while this is not a prevalent component in the training set. When comparing Fig. \ref{cdvaedist}b) and \ref{cdvaedist}e) and \ref{cdvaedist}c) and \ref{cdvaedist}f), we observe very different distributions for densities and packing fractions between the CDVAE structures and JARVIS training data. Specifically, the CDVAE generated structures have a much larger distribution of structures with a low density and packing fraction. This can be due in part to the low symmetry of the CDVAE generated structures, which gives rise to larger crystal volumes for the newly generated unit cells and therefore smaller densities and packing fractions. In fact, the CDVAE structures are all of space group P1. The likelihood of the CDVAE model generating low symmetry, chemically diverse structures might be attributed to the CDVAE model sampling from a Gaussian distribution to create new materials (to predict the number of atoms and composition). Since the underlying distribution of the materials is non-Gaussian, and the Gaussian distribution that the CDVAE model samples from is not representative of the latent space, materials out of the distribution can be generated. This is a limitation of CDVAE which could be addressed with a better latent space encoding in the future.

\begin{figure}[h]
    \centering
    \includegraphics[trim={0. 0cm 0 0cm},clip,width=0.7\textwidth]{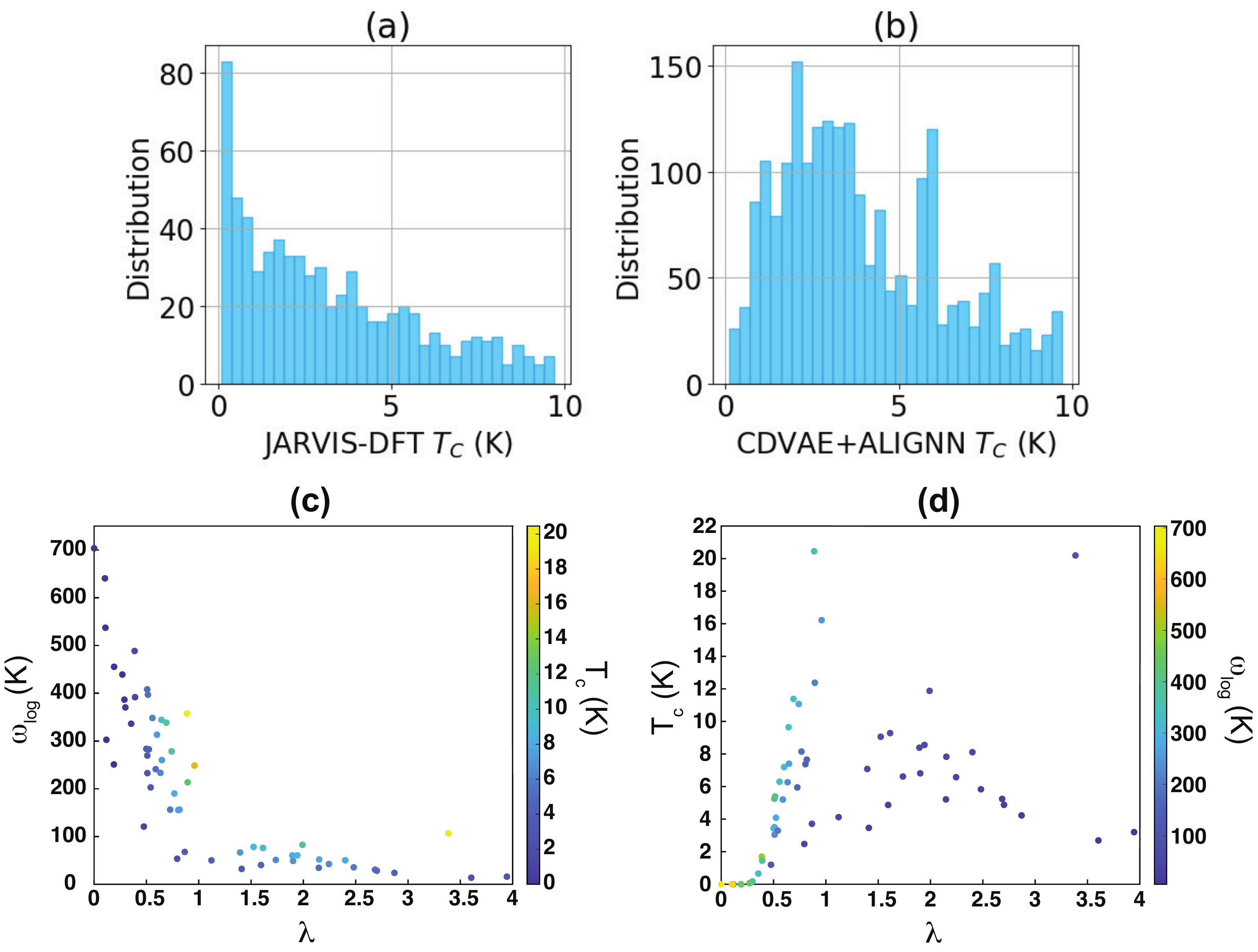}
    \caption{The distribution of $T_c$ for the a) 1058 JARVIS-DFT structures ($T_c$ computed with DFT) and b) the 3000 CDVAE structures ($T_c$ computed with ALIGNN). c) - d) the relation between EPC parameters for the CDVAE candidate materials verified with DFT.   }
    \label{cdvaefull}
\end{figure}

After discussing the structural and chemical diversity of the CDVAE generated structures, we plan to discuss the steps of the workflow that involve the prediction of $T_c$ with deep learning (ALIGNN) and DFT. The ALIGNN deep learning property predictor for $T_c$ was previously pretrained on the dataset in ref. \cite{bulksc} (1058 materials) and benchmarked. Using the pretrained ALIGNN model allows us to filter the vast amount of CDVAE candidate superconductors with instantaneous property prediction. Fig. \ref{cdvaefull}a) and b) depict the distribution of $T_c$ for the a) 1058 JARVIS-DFT structures ($T_c$ computed with DFT) and b) the 3000 CDVAE structures ($T_c$ computed with ALIGNN). As seen in the figure, the distribution of $T_c$ is vastly different. Most strikingly, the CDVAE generated data has a very low amount of structures with a predicted $T_c$ close to zero (non-superconducting), as opposed to a large number of materials in the JARVIS training set that have a $T_c$ close to zero. In addition, the distribution of CDVAE+ALIGNN temperatures resembles more of a Gaussian distribution, with a majority of the $T_c$ values of the newly generated structures concentrated around 2.5 K - 6 K. This can be an indicator of the success of the CDVAE model for generating new structures with an optimized target property, such as $T_c$, in the latent space. 

After screening the 3000 CDVAE generated structures with ALIGNN, we performed DFT for the 61 candidates that met the ALIGNN screening criteria depicted in Fig. \ref{cdvaeworkflow}a). At this stage, it is important to acknowledge the success of the pretrained ALIGNN model with regards to predicting the $T_c$ of these newly generated structures. Out of the 61 structures, 54 of them have a DFT computed $T_c$ $>$ 0.5 K, which indicates that ALIGNN is 89 $\%$ successful with filtering out superconducting materials from the CDVAE generated structures. Out of the 61 structures, 32 of them have a DFT computed $T_c$ $>$ 5 K, which indicates that ALIGNN has a 52 $\%$ success rate in filtering out materials with a $T_c$ above 5 K from the CDVAE generated structures. More details of the ALIGNN screening are given in the SI. It is important to note that the ALIGNN predictions used for prescreening were performed on structures generated from CDVAE prior to geometric relaxation with DFT. A full description of the DFT and ALIGNN (on the relaxed and unrelaxed structures) predictions are given in Table S1. To demonstrate how the predictive power of ALIGNN extrapolates from unrelaxed structures to relaxed structures for $T_c$ predictions, we plot the ALIGNN $T_c$ from the unrelaxed structures versus the ALIGNN $T_c$ from the relaxed structures in Fig. S2, where we observe that the predictions are relatively consistent between structures. We perform similar analysis for formation energy and band gap in Fig. S3 and Table S2. For $E_\textrm{form}$, we observe much closer agreement when comparing DFT and ALIGNN predictions for the relaxed (MAE of 0.15 eV/atom) structures compared to ALIGNN predictions for the unrelaxed (MAE of 0.42 eV/atom) structures (see Fig. S3). With regards to band gap screening, for 41 of the materials we ran DFT for, DFT and ALIGNN predicted them to be metallic (in relaxed and unrelaxed configurations). For the remaining 20 structures, at least one of the three methods predicted a band gap (either DFT, ALIGNN on unrelaxed structures or ALIGNN on relaxed structures). A comparison of these band gap results is given in Table S2. The success of the ALIGNN prediction of $T_c$ is a direct consequence of the availability of training data. In terms of deep learning models for material property prediction, 1058 structures is a relatively small amount of data to train on. The reason for the smaller training set size is due to the vast computational expense needed to perform these DFT calculations of EPC properties. In principle, the ALIGNN prediction of $T_c$ can be systematically improved by adding more DFT calculations to the training, which is an ongoing effort of JARVIS. We show the relationship between EPC parameters ($\lambda$, $\omega_{\textrm{log}}$, $T_c$) for the 61 materials we performed DFT calculations for in Fig. \ref{cdvaefull}c) and d). Fig. \ref{cdvaefull}c) depicts an inverse relationship between $\lambda$ and $\omega_{\textrm{log}}$ and in Fig. \ref{cdvaefull}d), we observe a somewhat positive relationship between $\lambda$ and $T_c$. These are typical behaviors of BCS superconductors and were observed in our work on BCS bulk and 2D superconductors \cite{bulksc,2dsc}. From the colormap of Fig. \ref{cdvaefull}c) and d), it is clear that a balance of high $\lambda$ and $\omega_{\textrm{log}}$ is a necessary condition for a material to have a high $T_c$. It is important to note that our focus on ambient condition stoichiometric BCS superconductors significantly limits the value of $T_c$ that can be achieved. High-pressure superconductors, non-stoichiometric (i.e. several cuprates and pnictides) superconductors, and unconventional superconductors (not mediated by electron-phonon interactions) can achieve much higher $T_c$ than BCS superconductors. We hope to extend our workflow to these types of superconductors in the future, but for proof-of-concept and validation of our methods, we focus on BCS superconductors in this work.

\begin{figure*}[h]
    \centering
    \includegraphics[trim={0. 0cm 0 0cm},clip,width=0.8\textwidth]{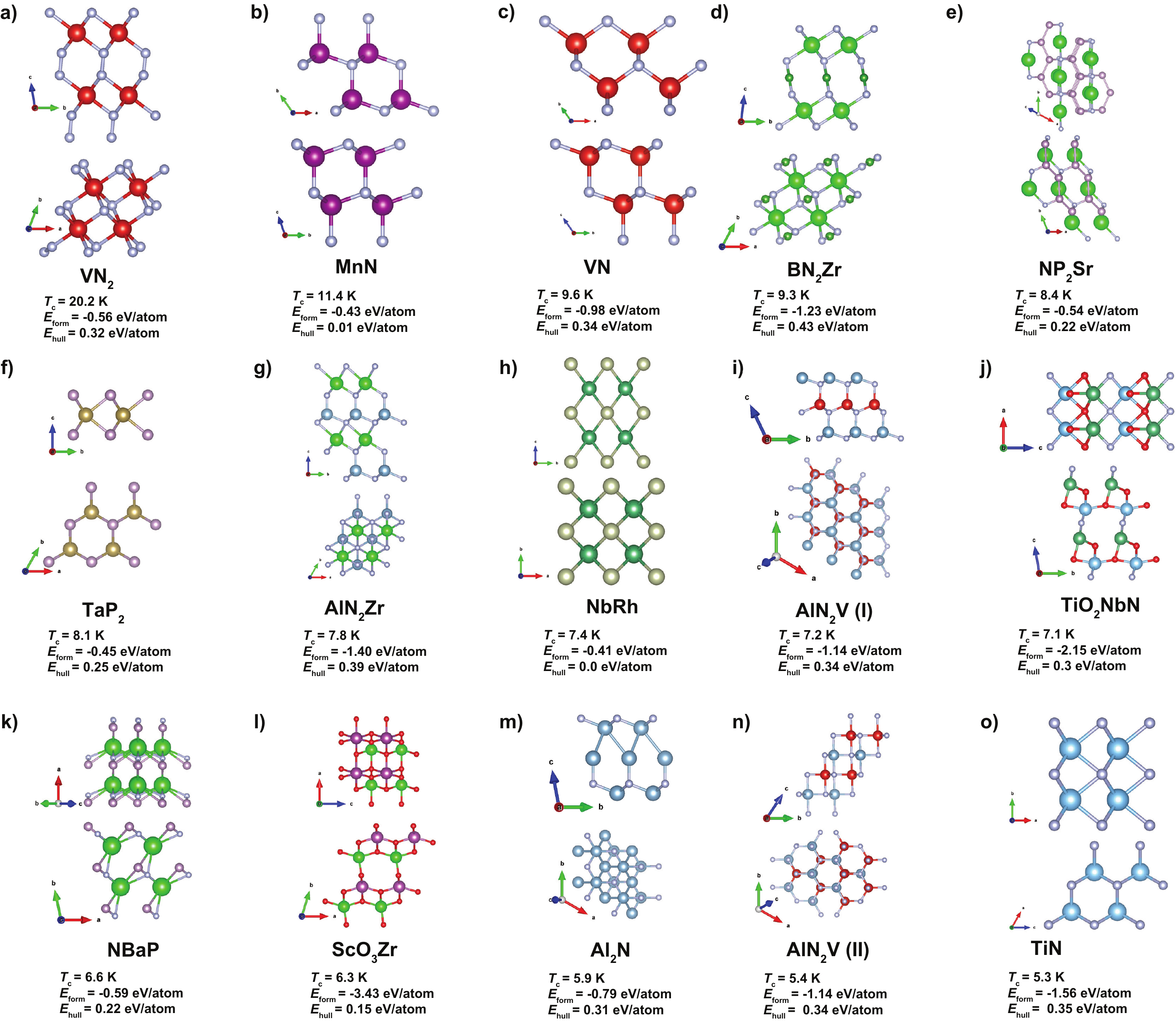}
    \caption{Top and side view of the top superconductor candidates (closest to the convex hull) generated with CDVAE and verified with DFT: a) VN$_2$, b) MnN, c) VN, d) BN$_2$Zr, e) NP$_2$Sr, f) TaP$_2$, g) AlN$_2$Zr, h) NbRh, i) AlN$_2$V (I), j) TiO$_2$NbN, k) NBaP, l) ScO$_3$Zr, m) Al$_2$N, n) AlN$_2$V (II), o) TiN. $T_c$, $E_\textrm{form}$ and $E_\textrm{hull}$ are also given for each material.   }
    \label{structures}
\end{figure*}

%nms=['POSCAR-VN2_PBEBO','POSCAR-NTaB_PBEBO','POSCAR-BORu_PBEBO','POSCAR-BTa2N_PBEBO','POSCAR-NTaB_PBEBO','POSCAR-BN2Zr_PBEBO','POSCAR-BTaNS_PBEBO','POSCAR-NP2Sr_PBEBO','POSCAR-TaP2_PBEBO','POSCAR-NPdTi2_PBEBO','POSCAR-PScSi2_PBEBO','POSCAR-AlN2V_PBEBO','POSCAR-TiO2NbN_PBEBO','POSCAR-NBaP_PBEBO','POSCAR-NVBRu_PBEBO','POSCAR-ScO3Zr_PBEBO','POSCAR-Al2N_PBEBO','POSCAR-AlN2V_PBEBO','POSCAR-TaONRu_PBEBO','POSCAR-ScBORuCa_PBEBO','POSCAR-B2TaS_PBEBO']

\begin{table}[t]
\caption{\label{topstructures} Chemical formula, JARVIS ID (JID), $T_c$, formation energy per atom and energy above the convex hull per atom of the 25candidate superconductors from CDVAE verified by DFT calculations.}
\label{table}
\begin{center}
\begin{tabular}{lllll}
\hline
 Structure  & JID &  $T_c$ & $E_\textrm{form}$   & $E_\textrm{hull}$\\
  & & (K)  & (eV/atom)  & (eV/atom)\\
\\ \hline \\
VN$_2$ &  JVASP-161655     & 20.2 & -0.56 & 0.32                 \\
NTaB (I) & JVASP-161630 & 16.2 & -0.32 & 0.84                 \\
BORu  &  JVASP-161610   & 12.4 & -0.72 & 0.75                 \\
BTa$_2$N & JVASP-161612    & 11.9 & -0.37 & 0.67                 \\
MnN	& JVASP-161621 & 11.4	& -0.43	& 0.01
\\
NTaB (II) & JVASP-161624 & 11.1 & -0.32 & 0.84 \\
VN	& JVASP-161656	& 9.6	& -0.98	& 0.34 \\
BN$_2$Zr  & JVASP-161608   & 9.3  & -1.23 & 0.43                 \\
BTaNS   &  JVASP-161613 & 9.1  & -0.33 & 0.88                 \\
NP$_2$Sr  & JVASP-162663    & 8.4  & -0.54 & 0.22 \\
TaP$_2$  & JVASP-161649    & 8.1  & -0.45 & 0.25                 \\
AlN$_2$Zr	& JVASP-161600	& 7.8	& -1.40	& 0.39 \\
NPdTi$_2$ & JVASP-161629   & 7.7  & -0.78 & 0.47                 \\
PScSi$_2$   &  JVASP-161644 & 7.4  & -0.43 & 0.49                 \\
NbRh	& JVASP-161638	& 7.4	& -0.41	& 0.00 \\
AlN$_2$V (I) & JVASP-161599 & 7.2  & -1.14 & 0.34                 \\
TiO$_2$NbN  & JVASP-161653   & 7.1  & -2.15 & 0.30                 \\
NBaP &   JVASP-161626   & 6.6  & -0.59 & 0.22                 \\
NVBRu  &  JVASP-161631  & 6.3  & -0.14 & 0.74                 \\
ScO$_3$Zr  & JVASP-161647  & 6.3  & -3.43 & 0.15                 \\
Al$_2$N   &  JVASP-161597  & 5.9  & -0.79 & 0.31                 \\
AlN$_2$V (II) & JVASP-162662 & 5.4  & -1.14 & 0.34 \\
TiN	& JVASP-161652 &	5.3	 & -1.56	& 0.35 \\
ScBORuCa & JVASP-161646 & 5.2  & -1.12 & 0.58                 \\
B$_2$TaS  & JVASP-161604   & 5.2  & -0.07 & 0.58    

\end{tabular}
\end{center}
\end{table}

The chemical composition, $T_c$, formation energy per atom and energy above the convex hull of the top superconducting candidates are given in Table \ref{topstructures}. As seen in the table, all of the 25 structures have negative formation energy. Although negative formation energy is a necessary prerequisite for thermodynamic stability, it does not guarantee thermodynamic stability. For this reason, we calculated the energy above the convex hull (shown in Table \ref{topstructures}). We computed the convex hull of these structures from the formation energy calculations of the different phases in JARVIS-DFT. From here, we observed 15 structures within an energy of 0.45 eV or less from the convex hull, and deem that these structures are the most likely to be experimentally synthesized. The atomic structures (all of P1 symmetry) of these 15 materials are given in Fig. \ref{structures} (VN$_2$, MnN, VN, BN$_2$Zr, NP$_2$Sr, TaP$_2$, AlN$_2$Zr, NbRh, AlN$_2$V (I), TiO$_2$NbN, NBaP, ScO$_3$Zr, Al$_2$N, AlN$_2$V (II), TiN). The properties and atomic structures of all the materials we performed DFT verification for (61 materials) will be available along with this manuscript and uploaded to the JARVIS-DFT database (\url{https://jarvis.nist.gov/}). Although these 15 materials have an energy above the convex hull between 0.0 eV/atom and 0.43 eV/atom, indicating that thermodynamic stability is still not guaranteed, they still may be synthesizable. In fact, Aykol et al. \cite{doi:10.1126/sciadv.aaq0148} demonstrated that nitride-based polymorphs with a higher energy above the convex hull can be stable up to a high energetic amorphous limit. Since a majority of these superconductors contain nitrogen, this gives promise that they can eventually be experimentally realized.

As a success metric of the CDVAE method for inverse design, it is important to measure how diverse the chemical composition and crystal structure of the generated materials are, not only with respect to the training data, but with respect to larger areas of phase space. Out of the top 25 candidate superconductors, 20 do not have a chemical composition in the JARVIS training set used for CDVAE. Compounds with a stoichiometry of MnN, VN, AlN$_2$Zr, NbRh and TiN are part of the training set and were all found to have superconducting phases in ref. \cite{bulksc}. Although these five candidates have similar chemistry to compounds in the training set, they all have entirely different crystal structures, with the exception of NbRh (which has the same structure as JVASP-20529). Due to the fact that the JARVIS database has less materials (the primary focus of JARVIS is to expand the properties and accuracy of materials), we checked other well-established materials databases for these candidates. Specifically, we checked the Materials Project (over 150000 structures) and OQMD (over 1000000 structures). In addition, we searched the literature to see if any of these materials have been synthesized. For VN$_2$, we found 3 compounds in OQMD with entirely different structures (1239835, 1233906, 1589664) with $E_\textrm{hull}$ ranging from 0.8 to 2.6 eV/atom. For Al$_2$N, we found 4 compounds in OQMD with entirely different structures (1237731, 1415673, 1590028) with $E_\textrm{hull}$ ranging from (0.9 to 2.6) eV/atom. We found 3 TaP$_2$ compounds (1372133, 1590238, 1237725) in OQMD. The $E_\textrm{hull}$ ranged from 0 to 1.2 eV/atom, but none of these phases resemble the structure in this study, which has a similar crystal structure to layered transition metal dichalcogenides (such as 2H-MoS$_2$). Upon searching the literature, we found that a semimetallic phase of TaP$_2$ had been recently synthesized \cite{PhysRevB.100.115127} that possessed a different crystal structure than the superconducting phase reported in this work. This experimentally synthesized TaP$_2$ was found to possess quantum oscillations and nontrivial topological properties, but was not found to be superconducting \cite{PhysRevB.100.115127}. We found 1 AlN$_2$V compound in the Materials Project (mp-1247742), where the $E_\textrm{hull}$ was 0.37 eV/atom. When comparing this structure to the two AlN$_2$V structures in our study, mp-1247742 has a similar form to AlN$_2$V (I) but a drastically different structure than AlN$_2$V (II). With regards to the five structures that have the same stoichiometry as structures in the training set, MnN, VN, NbRh and TiN have multiple entries in OQMD and Materials Project, while AlN$_2$Zr has just one entry in Materials Project. Our NbRh structure is the same as OQMD 30785, 17746, 339226, mp-1963 and JVASP-20529 (our $T_c$ is in agreement with the value obtained in ref. \cite{bulksc}) and our TiN structure is similar to the OQMD 1230416 entry. Upon searching the experimental literature, certain phases of TiN and VN have been reported to have superconducting properties \cite{92701,Romanov_2018,PhysRevApplied.12.054001,6942188}. We additionally searched the Supercon database (the Supercon database is unavailable as of December 2021, but previous papers have made this dataset available for further research \cite{supercon,hamidieh2018datadriven}), which consists of over 16000 experimentally realized superconducting materials. We find no common materials between the top 25 CDVAE candidates and the Supercon database. This further proves that the CDVAE method can generate unique materials with a specific desired property, covering previously undiscovered areas of phase space.

In this work, we used a multi step workflow, combining generative models, deep learning property prediction and DFT to discover next-generation superconducting materials. We demonstrated that the deep learning property prediction using the ALIGNN model can accelerate the search for new superconductors by instantaneously screening the properties of the newly generated materials prior to DFT verification and experimental investigation. Our search revealed 25 newly predicted candidate superconductors, with 15 structures being close to the convex hull and $T_c$ values as high as 20.2 K. Our approach goes beyond the standard funnel-like materials design workflow and allows for inverse design of novel materials, populating previously undiscovered areas of phase space. 

\section{Data Availability Statement}
The data from the present work is available at \url{https://figshare.com/articles/dataset/Inverse_Design_of_Next-Generation_Superconductors_Using_Data-Driven_Deep_Generative_Models/23681025}.

 \section{Code Availability Statement}
Software packages mentioned in the article can be found at \url{https://github.com/usnistgov/jarvis}, \url{https://github.com/knc6/cdvae} and \url{https://github.com/txie-93/cdvae}.

 \section{Competing interests}
The authors declare no competing interests.

 \section{Supporting Information}
CDVAE loss function, tabulated data for relaxed and unrelaxed structures computed with ALIGNN and DFT, additional comparisons of ALIGNN for relaxed and unrelaxed structures.
 
\section{Acknowledgments}
All authors thank the National Institute of Standards and Technology for funding, computational, and data-management resources, specifically the NIST Nisaba and Raritan HPC clusters. K.C. thanks the computational support from XSEDE (Extreme Science and Engineering Discovery Environment) computational resources under allocation number TG-DMR 190095. Contributions from K.C. were supported by the financial assistance award 70NANB19H117 from the U.S. Department of Commerce, National Institute of Standards and Technology.

\bibliography{main}

\providecommand{\noopsort}[1]{}\providecommand{\singleletter}[1]{#1}%
\providecommand{\latin}[1]{#1}
\makeatletter
\providecommand{\doi}
  {\begingroup\let\do\@makeother\dospecials
  \catcode`\{=1 \catcode`\}=2 \doi@aux}
\providecommand{\doi@aux}[1]{\endgroup\texttt{#1}}
\makeatother
\providecommand*\mcitethebibliography{\thebibliography}
\csname @ifundefined\endcsname{endmcitethebibliography}
  {\let\endmcitethebibliography\endthebibliography}{}
\begin{mcitethebibliography}{69}
\providecommand*\natexlab[1]{#1}
\providecommand*\mciteSetBstSublistMode[1]{}
\providecommand*\mciteSetBstMaxWidthForm[2]{}
\providecommand*\mciteBstWouldAddEndPuncttrue
  {\def\EndOfBibitem{\unskip.}}
\providecommand*\mciteBstWouldAddEndPunctfalse
  {\let\EndOfBibitem\relax}
\providecommand*\mciteSetBstMidEndSepPunct[3]{}
\providecommand*\mciteSetBstSublistLabelBeginEnd[3]{}
\providecommand*\EndOfBibitem{}
\mciteSetBstSublistMode{f}
\mciteSetBstMaxWidthForm{subitem}{(\alph{mcitesubitemcount})}
\mciteSetBstSublistLabelBeginEnd
  {\mcitemaxwidthsubitemform\space}
  {\relax}
  {\relax}

\bibitem[Kamerlingh~Onnes(1911)]{kamerlingh1911resistance}
Kamerlingh~Onnes,~H. The resistance of pure mercury at helium temperatures.
  \emph{Commun. Phys. Lab. Univ. Leiden, b} \textbf{1911}, \emph{120}\relax
\mciteBstWouldAddEndPuncttrue
\mciteSetBstMidEndSepPunct{\mcitedefaultmidpunct}
{\mcitedefaultendpunct}{\mcitedefaultseppunct}\relax
\EndOfBibitem
\bibitem[Poole \latin{et~al.}(2013)Poole, Farach, and
  Creswick]{poole2013superconductivity}
Poole,~C.~P.; Farach,~H.~A.; Creswick,~R.~J. \emph{Superconductivity}; Academic
  press, 2013\relax
\mciteBstWouldAddEndPuncttrue
\mciteSetBstMidEndSepPunct{\mcitedefaultmidpunct}
{\mcitedefaultendpunct}{\mcitedefaultseppunct}\relax
\EndOfBibitem
\bibitem[Rogalla and Kes(2011)Rogalla, and Kes]{rogalla2011100}
Rogalla,~H.; Kes,~P.~H. \emph{100 years of superconductivity}; Taylor \&
  Francis, 2011\relax
\mciteBstWouldAddEndPuncttrue
\mciteSetBstMidEndSepPunct{\mcitedefaultmidpunct}
{\mcitedefaultendpunct}{\mcitedefaultseppunct}\relax
\EndOfBibitem
\bibitem[Cooper and Feldman(2010)Cooper, and Feldman]{cooper2010bcs}
Cooper,~L.~N.; Feldman,~D. \emph{BCS: 50 years}; World scientific, 2010\relax
\mciteBstWouldAddEndPuncttrue
\mciteSetBstMidEndSepPunct{\mcitedefaultmidpunct}
{\mcitedefaultendpunct}{\mcitedefaultseppunct}\relax
\EndOfBibitem
\bibitem[Giustino(2017)]{giustino2017electron}
Giustino,~F. Electron-phonon interactions from first principles. \emph{Reviews
  of Modern Physics} \textbf{2017}, \emph{89}, 015003\relax
\mciteBstWouldAddEndPuncttrue
\mciteSetBstMidEndSepPunct{\mcitedefaultmidpunct}
{\mcitedefaultendpunct}{\mcitedefaultseppunct}\relax
\EndOfBibitem
\bibitem[Choudhary and Garrity(2022)Choudhary, and Garrity]{bulksc}
Choudhary,~K.; Garrity,~K. Designing high-TC superconductors with BCS-inspired
  screening, density functional theory, and deep-learning. \emph{npj
  Computational Materials} \textbf{2022}, \emph{8}, 244\relax
\mciteBstWouldAddEndPuncttrue
\mciteSetBstMidEndSepPunct{\mcitedefaultmidpunct}
{\mcitedefaultendpunct}{\mcitedefaultseppunct}\relax
\EndOfBibitem
\bibitem[Wines \latin{et~al.}(2023)Wines, Choudhary, Biacchi, Garrity, and
  Tavazza]{2dsc}
Wines,~D.; Choudhary,~K.; Biacchi,~A.~J.; Garrity,~K.~F.; Tavazza,~F.
  High-throughput DFT-based discovery of next generation two-dimensional (2D)
  superconductors. \emph{Nano Letters} \textbf{2023}, \emph{23}, 969--978\relax
\mciteBstWouldAddEndPuncttrue
\mciteSetBstMidEndSepPunct{\mcitedefaultmidpunct}
{\mcitedefaultendpunct}{\mcitedefaultseppunct}\relax
\EndOfBibitem
\bibitem[Shipley \latin{et~al.}(2021)Shipley, Hutcheon, Needs, and
  Pickard]{PhysRevB.104.054501}
Shipley,~A.~M.; Hutcheon,~M.~J.; Needs,~R.~J.; Pickard,~C.~J. High-throughput
  discovery of high-temperature conventional superconductors. \emph{Phys. Rev.
  B} \textbf{2021}, \emph{104}, 054501\relax
\mciteBstWouldAddEndPuncttrue
\mciteSetBstMidEndSepPunct{\mcitedefaultmidpunct}
{\mcitedefaultendpunct}{\mcitedefaultseppunct}\relax
\EndOfBibitem
\bibitem[Garc{\'\i}a-Nieto \latin{et~al.}(2021)Garc{\'\i}a-Nieto,
  Garc{\'\i}a-Gonzalo, and Paredes-S{\'a}nchez]{ml-sc1}
Garc{\'\i}a-Nieto,~P.~J.; Garc{\'\i}a-Gonzalo,~E.; Paredes-S{\'a}nchez,~J.
  Prediction of the critical temperature of a superconductor by using the
  WOA/MARS, Ridge, Lasso and Elastic-net machine learning techniques.
  \emph{Neural Computing and Applications} \textbf{2021}, \emph{33},
  17131--17145\relax
\mciteBstWouldAddEndPuncttrue
\mciteSetBstMidEndSepPunct{\mcitedefaultmidpunct}
{\mcitedefaultendpunct}{\mcitedefaultseppunct}\relax
\EndOfBibitem
\bibitem[Stanev \latin{et~al.}(2018)Stanev, Oses, Kusne, Rodriguez, Paglione,
  Curtarolo, and Takeuchi]{ml-sc2}
Stanev,~V.; Oses,~C.; Kusne,~A.~G.; Rodriguez,~E.; Paglione,~J.; Curtarolo,~S.;
  Takeuchi,~I. Machine learning modeling of superconducting critical
  temperature. \emph{npj Computational Materials} \textbf{2018}, \emph{4},
  29\relax
\mciteBstWouldAddEndPuncttrue
\mciteSetBstMidEndSepPunct{\mcitedefaultmidpunct}
{\mcitedefaultendpunct}{\mcitedefaultseppunct}\relax
\EndOfBibitem
\bibitem[Zhang \latin{et~al.}(2022)Zhang, Zhu, Xiang, Zhang, Huang, Zhong, Qiu,
  Hu, and Lin]{ml-sc3}
Zhang,~J.; Zhu,~Z.; Xiang,~X.~D.; Zhang,~K.; Huang,~S.; Zhong,~C.; Qiu,~H.-J.;
  Hu,~K.; Lin,~X. Machine learning prediction of superconducting critical
  temperature through the structural descriptor. \emph{The Journal of Physical
  Chemistry C} \textbf{2022}, \emph{126}, 8922--8927\relax
\mciteBstWouldAddEndPuncttrue
\mciteSetBstMidEndSepPunct{\mcitedefaultmidpunct}
{\mcitedefaultendpunct}{\mcitedefaultseppunct}\relax
\EndOfBibitem
\bibitem[Meredig \latin{et~al.}(2018)Meredig, Antono, Church, Hutchinson, Ling,
  Paradiso, Blaiszik, Foster, Gibbons, Hattrick-Simpers, Mehta, and
  Ward]{C8ME00012C}
Meredig,~B.; Antono,~E.; Church,~C.; Hutchinson,~M.; Ling,~J.; Paradiso,~S.;
  Blaiszik,~B.; Foster,~I.; Gibbons,~B.; Hattrick-Simpers,~J.; Mehta,~A.;
  Ward,~L. Can machine learning identify the next high-temperature
  superconductor? Examining extrapolation performance for materials discovery.
  \emph{Mol. Syst. Des. Eng.} \textbf{2018}, \emph{3}, 819--825\relax
\mciteBstWouldAddEndPuncttrue
\mciteSetBstMidEndSepPunct{\mcitedefaultmidpunct}
{\mcitedefaultendpunct}{\mcitedefaultseppunct}\relax
\EndOfBibitem
\bibitem[Roter and Dordevic(2020)Roter, and Dordevic]{ROTER20201353689}
Roter,~B.; Dordevic,~S. Predicting new superconductors and their critical
  temperatures using machine learning. \emph{Physica C: Superconductivity and
  its Applications} \textbf{2020}, \emph{575}, 1353689\relax
\mciteBstWouldAddEndPuncttrue
\mciteSetBstMidEndSepPunct{\mcitedefaultmidpunct}
{\mcitedefaultendpunct}{\mcitedefaultseppunct}\relax
\EndOfBibitem
\bibitem[Menon and Ranganathan(2022)Menon, and Ranganathan]{ml-sc4}
Menon,~D.; Ranganathan,~R. A generative approach to materials discovery,
  design, and optimization. \emph{ACS Omega} \textbf{2022}, \emph{7},
  25958--25973\relax
\mciteBstWouldAddEndPuncttrue
\mciteSetBstMidEndSepPunct{\mcitedefaultmidpunct}
{\mcitedefaultendpunct}{\mcitedefaultseppunct}\relax
\EndOfBibitem
\bibitem[Jain \latin{et~al.}(2013)Jain, Ong, Hautier, Chen, Richards, Dacek,
  Cholia, Gunter, Skinner, Ceder, and Persson]{doi:10.1063/1.4812323}
Jain,~A.; Ong,~S.~P.; Hautier,~G.; Chen,~W.; Richards,~W.~D.; Dacek,~S.;
  Cholia,~S.; Gunter,~D.; Skinner,~D.; Ceder,~G.; Persson,~K.~A. Commentary:
  The Materials Project: A materials genome approach to accelerating materials
  innovation. \emph{APL Materials} \textbf{2013}, \emph{1}, 011002\relax
\mciteBstWouldAddEndPuncttrue
\mciteSetBstMidEndSepPunct{\mcitedefaultmidpunct}
{\mcitedefaultendpunct}{\mcitedefaultseppunct}\relax
\EndOfBibitem
\bibitem[Saal \latin{et~al.}(2013)Saal, Kirklin, Aykol, Meredig, and
  Wolverton]{oqmd}
Saal,~J.~E.; Kirklin,~S.; Aykol,~M.; Meredig,~B.; Wolverton,~C. Materials
  design and discovery with high-throughput density functional theory: The Open
  Quantum Materials Database (OQMD). \emph{JOM} \textbf{2013}, \emph{65},
  1501--1509\relax
\mciteBstWouldAddEndPuncttrue
\mciteSetBstMidEndSepPunct{\mcitedefaultmidpunct}
{\mcitedefaultendpunct}{\mcitedefaultseppunct}\relax
\EndOfBibitem
\bibitem[Kirklin \latin{et~al.}(2015)Kirklin, Saal, Meredig, Thompson, Doak,
  Aykol, R{\"u}hl, and Wolverton]{oqmd-2}
Kirklin,~S.; Saal,~J.~E.; Meredig,~B.; Thompson,~A.; Doak,~J.~W.; Aykol,~M.;
  R{\"u}hl,~S.; Wolverton,~C. The Open Quantum Materials Database (OQMD):
  assessing the accuracy of DFT formation energies. \emph{npj Computational
  Materials} \textbf{2015}, \emph{1}, 15010\relax
\mciteBstWouldAddEndPuncttrue
\mciteSetBstMidEndSepPunct{\mcitedefaultmidpunct}
{\mcitedefaultendpunct}{\mcitedefaultseppunct}\relax
\EndOfBibitem
\bibitem[Haastrup \latin{et~al.}(2018)Haastrup, Strange, Pandey, Deilmann,
  Schmidt, Hinsche, Gjerding, Torelli, Larsen, Riis-Jensen, Gath, Jacobsen,
  Mortensen, Olsen, and Thygesen]{Haastrup_2018}
Haastrup,~S.; Strange,~M.; Pandey,~M.; Deilmann,~T.; Schmidt,~P.~S.;
  Hinsche,~N.~F.; Gjerding,~M.~N.; Torelli,~D.; Larsen,~P.~M.;
  Riis-Jensen,~A.~C.; Gath,~J.; Jacobsen,~K.~W.; Mortensen,~J.~J.; Olsen,~T.;
  Thygesen,~K.~S. The Computational 2D Materials Database: high-throughput
  modeling and discovery of atomically thin crystals. \emph{2D Materials}
  \textbf{2018}, \emph{5}, 042002\relax
\mciteBstWouldAddEndPuncttrue
\mciteSetBstMidEndSepPunct{\mcitedefaultmidpunct}
{\mcitedefaultendpunct}{\mcitedefaultseppunct}\relax
\EndOfBibitem
\bibitem[Gjerding \latin{et~al.}(2021)Gjerding, Taghizadeh, Rasmussen, Ali,
  Bertoldo, Deilmann, Kn{\o}sgaard, Kruse, Larsen, Manti, Pedersen, Petralanda,
  Skovhus, Svendsen, Mortensen, Olsen, and Thygesen]{Gjerding_2021}
Gjerding,~M.~N. \latin{et~al.}  Recent progress of the Computational 2D
  Materials Database (C2DB). \emph{2D Materials} \textbf{2021}, \emph{8},
  044002\relax
\mciteBstWouldAddEndPuncttrue
\mciteSetBstMidEndSepPunct{\mcitedefaultmidpunct}
{\mcitedefaultendpunct}{\mcitedefaultseppunct}\relax
\EndOfBibitem
\bibitem[Choudhary \latin{et~al.}(2020)Choudhary, Garrity, Reid, DeCost,
  Biacchi, Hight~Walker, Trautt, Hattrick-Simpers, Kusne, Centrone,
  \latin{et~al.} others]{choudhary2020joint}
Choudhary,~K.; Garrity,~K.~F.; Reid,~A.~C.; DeCost,~B.; Biacchi,~A.~J.;
  Hight~Walker,~A.~R.; Trautt,~Z.; Hattrick-Simpers,~J.; Kusne,~A.~G.;
  Centrone,~A., \latin{et~al.}  The joint automated repository for various
  integrated simulations (JARVIS) for data-driven materials design. \emph{npj
  Computational Materials} \textbf{2020}, \emph{6}, 1--13\relax
\mciteBstWouldAddEndPuncttrue
\mciteSetBstMidEndSepPunct{\mcitedefaultmidpunct}
{\mcitedefaultendpunct}{\mcitedefaultseppunct}\relax
\EndOfBibitem
\bibitem[Ong \latin{et~al.}(2013)Ong, Richards, Jain, Hautier, Kocher, Cholia,
  Gunter, Chevrier, Persson, and Ceder]{ONG2013314}
Ong,~S.~P.; Richards,~W.~D.; Jain,~A.; Hautier,~G.; Kocher,~M.; Cholia,~S.;
  Gunter,~D.; Chevrier,~V.~L.; Persson,~K.~A.; Ceder,~G. Python Materials
  Genomics (pymatgen): A robust, open-source python library for materials
  analysis. \emph{Computational Materials Science} \textbf{2013}, \emph{68},
  314--319\relax
\mciteBstWouldAddEndPuncttrue
\mciteSetBstMidEndSepPunct{\mcitedefaultmidpunct}
{\mcitedefaultendpunct}{\mcitedefaultseppunct}\relax
\EndOfBibitem
\bibitem[Ramesh \latin{et~al.}(2021)Ramesh, Pavlov, Goh, Gray, Voss, Radford,
  Chen, and Sutskever]{ramesh2021zeroshot}
Ramesh,~A.; Pavlov,~M.; Goh,~G.; Gray,~S.; Voss,~C.; Radford,~A.; Chen,~M.;
  Sutskever,~I. Zero-shot text-to-image generation. 2021\relax
\mciteBstWouldAddEndPuncttrue
\mciteSetBstMidEndSepPunct{\mcitedefaultmidpunct}
{\mcitedefaultendpunct}{\mcitedefaultseppunct}\relax
\EndOfBibitem
\bibitem[Wu \latin{et~al.}(2022)Wu, Yang, van~den Berg, Zou, Lu, and
  Amini]{wu2022protein}
Wu,~K.~E.; Yang,~K.~K.; van~den Berg,~R.; Zou,~J.~Y.; Lu,~A.~X.; Amini,~A.~P.
  Protein structure generation via folding diffusion. 2022\relax
\mciteBstWouldAddEndPuncttrue
\mciteSetBstMidEndSepPunct{\mcitedefaultmidpunct}
{\mcitedefaultendpunct}{\mcitedefaultseppunct}\relax
\EndOfBibitem
\bibitem[Xie \latin{et~al.}(2021)Xie, Fu, Ganea, Barzilay, and
  Jaakkola]{https://doi.org/10.48550/arxiv.2110.06197}
Xie,~T.; Fu,~X.; Ganea,~O.-E.; Barzilay,~R.; Jaakkola,~T. Crystal diffusion
  variational autoencoder for periodic material generation. 2021;
  \url{https://arxiv.org/abs/2110.06197}\relax
\mciteBstWouldAddEndPuncttrue
\mciteSetBstMidEndSepPunct{\mcitedefaultmidpunct}
{\mcitedefaultendpunct}{\mcitedefaultseppunct}\relax
\EndOfBibitem
\bibitem[Lyngby and Thygesen(2022)Lyngby, and Thygesen]{cdvae-2d}
Lyngby,~P.; Thygesen,~K.~S. Data-driven discovery of 2D materials by deep
  generative models. \emph{npj Computational Materials} \textbf{2022},
  \emph{8}, 232\relax
\mciteBstWouldAddEndPuncttrue
\mciteSetBstMidEndSepPunct{\mcitedefaultmidpunct}
{\mcitedefaultendpunct}{\mcitedefaultseppunct}\relax
\EndOfBibitem
\bibitem[Moustafa \latin{et~al.}(2023)Moustafa, Lyngby, Mortensen, Thygesen,
  and Jacobsen]{PhysRevMaterials.7.014007}
Moustafa,~H.; Lyngby,~P.~M.; Mortensen,~J.~J.; Thygesen,~K.~S.; Jacobsen,~K.~W.
  Hundreds of new, stable, one-dimensional materials from a generative machine
  learning model. \emph{Phys. Rev. Mater.} \textbf{2023}, \emph{7},
  014007\relax
\mciteBstWouldAddEndPuncttrue
\mciteSetBstMidEndSepPunct{\mcitedefaultmidpunct}
{\mcitedefaultendpunct}{\mcitedefaultseppunct}\relax
\EndOfBibitem
\bibitem[Fung \latin{et~al.}(2022)Fung, Jia, Zhang, Bi, Yin, and
  Ganesh]{Fung_2022}
Fung,~V.; Jia,~S.; Zhang,~J.; Bi,~S.; Yin,~J.; Ganesh,~P. Atomic structure
  generation from reconstructing structural fingerprints. \emph{Machine
  Learning: Science and Technology} \textbf{2022}, \emph{3}, 045018\relax
\mciteBstWouldAddEndPuncttrue
\mciteSetBstMidEndSepPunct{\mcitedefaultmidpunct}
{\mcitedefaultendpunct}{\mcitedefaultseppunct}\relax
\EndOfBibitem
\bibitem[Noh \latin{et~al.}(2020)Noh, Gu, Kim, and Jung]{D0SC00594K}
Noh,~J.; Gu,~G.~H.; Kim,~S.; Jung,~Y. Machine-enabled inverse design of
  inorganic solid materials: promises and challenges. \emph{Chem. Sci.}
  \textbf{2020}, \emph{11}, 4871--4881\relax
\mciteBstWouldAddEndPuncttrue
\mciteSetBstMidEndSepPunct{\mcitedefaultmidpunct}
{\mcitedefaultendpunct}{\mcitedefaultseppunct}\relax
\EndOfBibitem
\bibitem[Kim \latin{et~al.}(2020)Kim, Noh, Gu, Aspuru-Guzik, and Jung]{genmod}
Kim,~S.; Noh,~J.; Gu,~G.~H.; Aspuru-Guzik,~A.; Jung,~Y. Generative adversarial
  networks for crystal structure prediction. \emph{ACS Central Science}
  \textbf{2020}, \emph{6}, 1412--1420\relax
\mciteBstWouldAddEndPuncttrue
\mciteSetBstMidEndSepPunct{\mcitedefaultmidpunct}
{\mcitedefaultendpunct}{\mcitedefaultseppunct}\relax
\EndOfBibitem
\bibitem[Long \latin{et~al.}(2021)Long, Fortunato, Opahle, Zhang, Samathrakis,
  Shen, Gutfleisch, and Zhang]{genmod2}
Long,~T.; Fortunato,~N.~M.; Opahle,~I.; Zhang,~Y.; Samathrakis,~I.; Shen,~C.;
  Gutfleisch,~O.; Zhang,~H. Constrained crystals deep convolutional generative
  adversarial network for the inverse design of crystal structures. \emph{npj
  Computational Materials} \textbf{2021}, \emph{7}, 66\relax
\mciteBstWouldAddEndPuncttrue
\mciteSetBstMidEndSepPunct{\mcitedefaultmidpunct}
{\mcitedefaultendpunct}{\mcitedefaultseppunct}\relax
\EndOfBibitem
\bibitem[Song \latin{et~al.}(2021)Song, Siriwardane, Zhao, and Hu]{genmod3}
Song,~Y.; Siriwardane,~E. M.~D.; Zhao,~Y.; Hu,~J. Computational discovery of
  new 2D materials using deep learning generative models. \emph{ACS Applied
  Materials \& Interfaces} \textbf{2021}, \emph{13}, 53303--53313\relax
\mciteBstWouldAddEndPuncttrue
\mciteSetBstMidEndSepPunct{\mcitedefaultmidpunct}
{\mcitedefaultendpunct}{\mcitedefaultseppunct}\relax
\EndOfBibitem
\bibitem[Noh \latin{et~al.}(2019)Noh, Kim, Stein, Sanchez-Lengeling, Gregoire,
  Aspuru-Guzik, and Jung]{NOH20191370}
Noh,~J.; Kim,~J.; Stein,~H.~S.; Sanchez-Lengeling,~B.; Gregoire,~J.~M.;
  Aspuru-Guzik,~A.; Jung,~Y. Inverse design of solid-state materials via a
  continuous representation. \emph{Matter} \textbf{2019}, \emph{1},
  1370--1384\relax
\mciteBstWouldAddEndPuncttrue
\mciteSetBstMidEndSepPunct{\mcitedefaultmidpunct}
{\mcitedefaultendpunct}{\mcitedefaultseppunct}\relax
\EndOfBibitem
\bibitem[Zhao \latin{et~al.}(2021)Zhao, Al-Fahdi, Hu, Siriwardane, Song,
  Nasiri, and Hu]{https://doi.org/10.1002/advs.202100566}
Zhao,~Y.; Al-Fahdi,~M.; Hu,~M.; Siriwardane,~E. M.~D.; Song,~Y.; Nasiri,~A.;
  Hu,~J. High-throughput discovery of novel cubic crystal materials using deep
  generative neural networks. \emph{Advanced Science} \textbf{2021}, \emph{8},
  2100566\relax
\mciteBstWouldAddEndPuncttrue
\mciteSetBstMidEndSepPunct{\mcitedefaultmidpunct}
{\mcitedefaultendpunct}{\mcitedefaultseppunct}\relax
\EndOfBibitem
\bibitem[Ren \latin{et~al.}(2022)Ren, Tian, Noh, Oviedo, Xing, Li, Liang, Zhu,
  Aberle, Sun, Wang, Liu, Li, Jayavelu, Hippalgaonkar, Jung, and
  Buonassisi]{REN2022314}
Ren,~Z. \latin{et~al.}  An invertible crystallographic representation for
  general inverse design of inorganic crystals with targeted properties.
  \emph{Matter} \textbf{2022}, \emph{5}, 314--335\relax
\mciteBstWouldAddEndPuncttrue
\mciteSetBstMidEndSepPunct{\mcitedefaultmidpunct}
{\mcitedefaultendpunct}{\mcitedefaultseppunct}\relax
\EndOfBibitem
\bibitem[Kingma and Welling(2013)Kingma, and
  Welling]{https://doi.org/10.48550/arxiv.1312.6114}
Kingma,~D.~P.; Welling,~M. Auto-encoding variational bayes. 2013;
  \url{https://arxiv.org/abs/1312.6114}\relax
\mciteBstWouldAddEndPuncttrue
\mciteSetBstMidEndSepPunct{\mcitedefaultmidpunct}
{\mcitedefaultendpunct}{\mcitedefaultseppunct}\relax
\EndOfBibitem
\bibitem[Sohl-Dickstein \latin{et~al.}(2015)Sohl-Dickstein, Weiss,
  Maheswaranathan, and Ganguli]{pmlr-v37-sohl-dickstein15}
Sohl-Dickstein,~J.; Weiss,~E.; Maheswaranathan,~N.; Ganguli,~S. Deep
  unsupervised learning using nonequilibrium thermodynamics. Proceedings of the
  32nd International Conference on Machine Learning. Lille, France, 2015; pp
  2256--2265\relax
\mciteBstWouldAddEndPuncttrue
\mciteSetBstMidEndSepPunct{\mcitedefaultmidpunct}
{\mcitedefaultendpunct}{\mcitedefaultseppunct}\relax
\EndOfBibitem
\bibitem[Song and Ermon(2019)Song, and
  Ermon]{https://doi.org/10.48550/arxiv.1907.05600}
Song,~Y.; Ermon,~S. Generative modeling by estimating gradients of the data
  distribution. 2019; \url{https://arxiv.org/abs/1907.05600}\relax
\mciteBstWouldAddEndPuncttrue
\mciteSetBstMidEndSepPunct{\mcitedefaultmidpunct}
{\mcitedefaultendpunct}{\mcitedefaultseppunct}\relax
\EndOfBibitem
\bibitem[Sohl-Dickstein \latin{et~al.}(2015)Sohl-Dickstein, Weiss,
  Maheswaranathan, and Ganguli]{sohldickstein2015deep}
Sohl-Dickstein,~J.; Weiss,~E.~A.; Maheswaranathan,~N.; Ganguli,~S. Deep
  unsupervised learning using nonequilibrium thermodynamics. 2015\relax
\mciteBstWouldAddEndPuncttrue
\mciteSetBstMidEndSepPunct{\mcitedefaultmidpunct}
{\mcitedefaultendpunct}{\mcitedefaultseppunct}\relax
\EndOfBibitem
\bibitem[Batzner \latin{et~al.}(2022)Batzner, Musaelian, Sun, Geiger, Mailoa,
  Kornbluth, Molinari, Smidt, and Kozinsky]{eq-gnn}
Batzner,~S.; Musaelian,~A.; Sun,~L.; Geiger,~M.; Mailoa,~J.~P.; Kornbluth,~M.;
  Molinari,~N.; Smidt,~T.~E.; Kozinsky,~B. E(3)-equivariant graph neural
  networks for data-efficient and accurate interatomic potentials. \emph{Nature
  Communications} \textbf{2022}, \emph{13}, 2453\relax
\mciteBstWouldAddEndPuncttrue
\mciteSetBstMidEndSepPunct{\mcitedefaultmidpunct}
{\mcitedefaultendpunct}{\mcitedefaultseppunct}\relax
\EndOfBibitem
\bibitem[Choudhary and DeCost(2021)Choudhary, and
  DeCost]{choudhary2021atomistic}
Choudhary,~K.; DeCost,~B. Atomistic Line Graph Neural Network for improved
  materials property predictions. \emph{npj Computational Materials}
  \textbf{2021}, \emph{7}, 1--8\relax
\mciteBstWouldAddEndPuncttrue
\mciteSetBstMidEndSepPunct{\mcitedefaultmidpunct}
{\mcitedefaultendpunct}{\mcitedefaultseppunct}\relax
\EndOfBibitem
\bibitem[Wang \latin{et~al.}(2019)Wang, Yu, Zheng, Gan, Gai, Ye, Li, Zhou,
  Huang, Ma, \latin{et~al.} others]{wang2019deep}
Wang,~M.; Yu,~L.; Zheng,~D.; Gan,~Q.; Gai,~Y.; Ye,~Z.; Li,~M.; Zhou,~J.;
  Huang,~Q.; Ma,~C., \latin{et~al.}  Deep Graph Library: Towards efficient and
  scalable deep learning on graphs. \emph{arXiv e-prints} \textbf{2019}, \relax
\mciteBstWouldAddEndPunctfalse
\mciteSetBstMidEndSepPunct{\mcitedefaultmidpunct}
{}{\mcitedefaultseppunct}\relax
\EndOfBibitem
\bibitem[Paszke \latin{et~al.}(2019)Paszke, Gross, Massa, Lerer, Bradbury,
  Chanan, Killeen, Lin, Gimelshein, Antiga, \latin{et~al.}
  others]{paszke2019pytorch}
Paszke,~A.; Gross,~S.; Massa,~F.; Lerer,~A.; Bradbury,~J.; Chanan,~G.;
  Killeen,~T.; Lin,~Z.; Gimelshein,~N.; Antiga,~L., \latin{et~al.}  Pytorch: An
  imperative style, high-performance deep learning library. \emph{Advances in
  neural information processing systems} \textbf{2019}, \emph{32},
  8026--8037\relax
\mciteBstWouldAddEndPuncttrue
\mciteSetBstMidEndSepPunct{\mcitedefaultmidpunct}
{\mcitedefaultendpunct}{\mcitedefaultseppunct}\relax
\EndOfBibitem
\bibitem[Choudhary \latin{et~al.}(2018)Choudhary, Zhang, Reid, Chowdhury,
  Van~Nguyen, Trautt, Newrock, Congo, and Tavazza]{choudhary2018computational}
Choudhary,~K.; Zhang,~Q.; Reid,~A.~C.; Chowdhury,~S.; Van~Nguyen,~N.;
  Trautt,~Z.; Newrock,~M.~W.; Congo,~F.~Y.; Tavazza,~F. Computational screening
  of high-performance optoelectronic materials using OptB88vdW and TB-mBJ
  formalisms. \emph{Scientific data} \textbf{2018}, \emph{5}, 1--12\relax
\mciteBstWouldAddEndPuncttrue
\mciteSetBstMidEndSepPunct{\mcitedefaultmidpunct}
{\mcitedefaultendpunct}{\mcitedefaultseppunct}\relax
\EndOfBibitem
\bibitem[Choudhary \latin{et~al.}(2019)Choudhary, Bercx, Jiang, Pachter,
  Lamoen, and Tavazza]{choudhary2019accelerated}
Choudhary,~K.; Bercx,~M.; Jiang,~J.; Pachter,~R.; Lamoen,~D.; Tavazza,~F.
  Accelerated discovery of efficient solar cell materials using quantum and
  machine-learning methods. \emph{Chemistry of Materials} \textbf{2019},
  \emph{31}, 5900--5908\relax
\mciteBstWouldAddEndPuncttrue
\mciteSetBstMidEndSepPunct{\mcitedefaultmidpunct}
{\mcitedefaultendpunct}{\mcitedefaultseppunct}\relax
\EndOfBibitem
\bibitem[Choudhary \latin{et~al.}(2021)Choudhary, Garrity, Ghimire, Anand, and
  Tavazza]{choudhary2021high}
Choudhary,~K.; Garrity,~K.~F.; Ghimire,~N.~J.; Anand,~N.; Tavazza,~F.
  High-throughput search for magnetic topological materials using spin-orbit
  spillage, machine learning, and experiments. \emph{Physical Review B}
  \textbf{2021}, \emph{103}, 155131\relax
\mciteBstWouldAddEndPuncttrue
\mciteSetBstMidEndSepPunct{\mcitedefaultmidpunct}
{\mcitedefaultendpunct}{\mcitedefaultseppunct}\relax
\EndOfBibitem
\bibitem[Choudhary \latin{et~al.}(2019)Choudhary, Garrity, and
  Tavazza]{choudhary2019high}
Choudhary,~K.; Garrity,~K.~F.; Tavazza,~F. High-throughput discovery of
  topologically non-trivial materials using spin-orbit spillage.
  \emph{Scientific reports} \textbf{2019}, \emph{9}, 1--8\relax
\mciteBstWouldAddEndPuncttrue
\mciteSetBstMidEndSepPunct{\mcitedefaultmidpunct}
{\mcitedefaultendpunct}{\mcitedefaultseppunct}\relax
\EndOfBibitem
\bibitem[Choudhary \latin{et~al.}(2020)Choudhary, Garrity, Jiang, Pachter, and
  Tavazza]{choudhary2020computational}
Choudhary,~K.; Garrity,~K.~F.; Jiang,~J.; Pachter,~R.; Tavazza,~F.
  Computational search for magnetic and non-magnetic 2D topological materials
  using unified spin--orbit spillage screening. \emph{NPJ Computational
  Materials} \textbf{2020}, \emph{6}, 1--8\relax
\mciteBstWouldAddEndPuncttrue
\mciteSetBstMidEndSepPunct{\mcitedefaultmidpunct}
{\mcitedefaultendpunct}{\mcitedefaultseppunct}\relax
\EndOfBibitem
\bibitem[Choudhary \latin{et~al.}(2018)Choudhary, Cheon, Reed, and
  Tavazza]{choudhary2018elastic}
Choudhary,~K.; Cheon,~G.; Reed,~E.; Tavazza,~F. Elastic properties of bulk and
  low-dimensional materials using van der Waals density functional.
  \emph{Physical Review B} \textbf{2018}, \emph{98}, 014107\relax
\mciteBstWouldAddEndPuncttrue
\mciteSetBstMidEndSepPunct{\mcitedefaultmidpunct}
{\mcitedefaultendpunct}{\mcitedefaultseppunct}\relax
\EndOfBibitem
\bibitem[Choudhary \latin{et~al.}(2020)Choudhary, Garrity, Sharma, Biacchi,
  Hight~Walker, and Tavazza]{choudhary2020high}
Choudhary,~K.; Garrity,~K.~F.; Sharma,~V.; Biacchi,~A.~J.; Hight~Walker,~A.~R.;
  Tavazza,~F. High-throughput density functional perturbation theory and
  machine learning predictions of infrared, piezoelectric, and dielectric
  responses. \emph{NPJ Computational Materials} \textbf{2020}, \emph{6},
  1--13\relax
\mciteBstWouldAddEndPuncttrue
\mciteSetBstMidEndSepPunct{\mcitedefaultmidpunct}
{\mcitedefaultendpunct}{\mcitedefaultseppunct}\relax
\EndOfBibitem
\bibitem[Choudhary \latin{et~al.}(2020)Choudhary, Ansari, Mazin, and
  Sauer]{choudhary2020density}
Choudhary,~K.; Ansari,~J.~N.; Mazin,~I.~I.; Sauer,~K.~L. Density functional
  theory-based electric field gradient database. \emph{Scientific Data}
  \textbf{2020}, \emph{7}, 1--10\relax
\mciteBstWouldAddEndPuncttrue
\mciteSetBstMidEndSepPunct{\mcitedefaultmidpunct}
{\mcitedefaultendpunct}{\mcitedefaultseppunct}\relax
\EndOfBibitem
\bibitem[Choudhary \latin{et~al.}(2017)Choudhary, Kalish, Beams, and
  Tavazza]{choudhary2017high}
Choudhary,~K.; Kalish,~I.; Beams,~R.; Tavazza,~F. High-throughput
  identification and characterization of two-dimensional materials using
  density functional theory. \emph{Scientific reports} \textbf{2017}, \emph{7},
  1--16\relax
\mciteBstWouldAddEndPuncttrue
\mciteSetBstMidEndSepPunct{\mcitedefaultmidpunct}
{\mcitedefaultendpunct}{\mcitedefaultseppunct}\relax
\EndOfBibitem
\bibitem[Wines \latin{et~al.}(2023)Wines, Choudhary, and Tavazza]{crx3-qmc}
Wines,~D.; Choudhary,~K.; Tavazza,~F. Systematic DFT+U and Quantum Monte Carlo
  benchmark of magnetic two-dimensional (2D) CrX3 (X = I, Br, Cl, F). \emph{The
  Journal of Physical Chemistry C} \textbf{2023}, \emph{127}, 1176--1188\relax
\mciteBstWouldAddEndPuncttrue
\mciteSetBstMidEndSepPunct{\mcitedefaultmidpunct}
{\mcitedefaultendpunct}{\mcitedefaultseppunct}\relax
\EndOfBibitem
\bibitem[Choudhary and Tavazza(2019)Choudhary, and
  Tavazza]{choudhary2019convergence}
Choudhary,~K.; Tavazza,~F. Convergence and machine learning predictions of
  Monkhorst-Pack k-points and plane-wave cut-off in high-throughput DFT
  calculations. \emph{Computational materials science} \textbf{2019},
  \emph{161}, 300--308\relax
\mciteBstWouldAddEndPuncttrue
\mciteSetBstMidEndSepPunct{\mcitedefaultmidpunct}
{\mcitedefaultendpunct}{\mcitedefaultseppunct}\relax
\EndOfBibitem
\bibitem[Baroni \latin{et~al.}(1987)Baroni, Giannozzi, and
  Testa]{baroni1987green}
Baroni,~S.; Giannozzi,~P.; Testa,~A. Green’s-function approach to linear
  response in solids. \emph{Physical review letters} \textbf{1987}, \emph{58},
  1861\relax
\mciteBstWouldAddEndPuncttrue
\mciteSetBstMidEndSepPunct{\mcitedefaultmidpunct}
{\mcitedefaultendpunct}{\mcitedefaultseppunct}\relax
\EndOfBibitem
\bibitem[Gonze(1995)]{gonze1995perturbation}
Gonze,~X. Perturbation expansion of variational principles at arbitrary order.
  \emph{Physical Review A} \textbf{1995}, \emph{52}, 1086\relax
\mciteBstWouldAddEndPuncttrue
\mciteSetBstMidEndSepPunct{\mcitedefaultmidpunct}
{\mcitedefaultendpunct}{\mcitedefaultseppunct}\relax
\EndOfBibitem
\bibitem[Wierzbowska \latin{et~al.}(2005)Wierzbowska, de~Gironcoli, and
  Giannozzi]{wierzbowska2005origins}
Wierzbowska,~M.; de~Gironcoli,~S.; Giannozzi,~P. Origins of low-and
  high-pressure discontinuities of $ T\_ $\{$c$\}$ $ in niobium. \emph{arXiv
  preprint cond-mat/0504077} \textbf{2005}, \relax
\mciteBstWouldAddEndPunctfalse
\mciteSetBstMidEndSepPunct{\mcitedefaultmidpunct}
{}{\mcitedefaultseppunct}\relax
\EndOfBibitem
\bibitem[Giannozzi \latin{et~al.}(2009)Giannozzi, Baroni, Bonini, Calandra,
  Car, Cavazzoni, Ceresoli, Chiarotti, Cococcioni, Dabo, \latin{et~al.}
  others]{giannozzi2009quantum}
Giannozzi,~P.; Baroni,~S.; Bonini,~N.; Calandra,~M.; Car,~R.; Cavazzoni,~C.;
  Ceresoli,~D.; Chiarotti,~G.~L.; Cococcioni,~M.; Dabo,~I., \latin{et~al.}
  QUANTUM ESPRESSO: a modular and open-source software project for quantum
  simulations of materials. \emph{Journal of physics: Condensed matter}
  \textbf{2009}, \emph{21}, 395502\relax
\mciteBstWouldAddEndPuncttrue
\mciteSetBstMidEndSepPunct{\mcitedefaultmidpunct}
{\mcitedefaultendpunct}{\mcitedefaultseppunct}\relax
\EndOfBibitem
\bibitem[Perdew \latin{et~al.}(2008)Perdew, Ruzsinszky, Csonka, Vydrov,
  Scuseria, Constantin, Zhou, and Burke]{perdew2008restoring}
Perdew,~J.~P.; Ruzsinszky,~A.; Csonka,~G.~I.; Vydrov,~O.~A.; Scuseria,~G.~E.;
  Constantin,~L.~A.; Zhou,~X.; Burke,~K. Restoring the density-gradient
  expansion for exchange in solids and surfaces. \emph{Physical review letters}
  \textbf{2008}, \emph{100}, 136406\relax
\mciteBstWouldAddEndPuncttrue
\mciteSetBstMidEndSepPunct{\mcitedefaultmidpunct}
{\mcitedefaultendpunct}{\mcitedefaultseppunct}\relax
\EndOfBibitem
\bibitem[Garrity \latin{et~al.}(2014)Garrity, Bennett, Rabe, and
  Vanderbilt]{garrity2014pseudopotentials}
Garrity,~K.~F.; Bennett,~J.~W.; Rabe,~K.~M.; Vanderbilt,~D. Pseudopotentials
  for high-throughput DFT calculations. \emph{Computational Materials Science}
  \textbf{2014}, \emph{81}, 446--452\relax
\mciteBstWouldAddEndPuncttrue
\mciteSetBstMidEndSepPunct{\mcitedefaultmidpunct}
{\mcitedefaultendpunct}{\mcitedefaultseppunct}\relax
\EndOfBibitem
\bibitem[McMillan(1968)]{mcmillan1968transition}
McMillan,~W. Transition temperature of strong-coupled superconductors.
  \emph{Physical Review} \textbf{1968}, \emph{167}, 331\relax
\mciteBstWouldAddEndPuncttrue
\mciteSetBstMidEndSepPunct{\mcitedefaultmidpunct}
{\mcitedefaultendpunct}{\mcitedefaultseppunct}\relax
\EndOfBibitem
\bibitem[Aykol \latin{et~al.}(2018)Aykol, Dwaraknath, Sun, and
  Persson]{doi:10.1126/sciadv.aaq0148}
Aykol,~M.; Dwaraknath,~S.~S.; Sun,~W.; Persson,~K.~A. Thermodynamic limit for
  synthesis of metastable inorganic materials. \emph{Science Advances}
  \textbf{2018}, \emph{4}, eaaq0148\relax
\mciteBstWouldAddEndPuncttrue
\mciteSetBstMidEndSepPunct{\mcitedefaultmidpunct}
{\mcitedefaultendpunct}{\mcitedefaultseppunct}\relax
\EndOfBibitem
\bibitem[Wang \latin{et~al.}(2019)Wang, Su, Zhang, Xia, Lin, Liu, Hou, Yu, Yu,
  Wang, Zou, Wang, Liang, Zhen, and Guo]{PhysRevB.100.115127}
Wang,~H.; Su,~H.; Zhang,~J.; Xia,~W.; Lin,~Y.; Liu,~X.; Hou,~X.; Yu,~Z.;
  Yu,~N.; Wang,~X.; Zou,~Z.; Wang,~Y.; Liang,~Q.; Zhen,~Y.; Guo,~Y. Quantum
  oscillations and nontrivial topological state in a compensated semimetal
  ${\mathrm{TaP}}_{2}$. \emph{Phys. Rev. B} \textbf{2019}, \emph{100},
  115127\relax
\mciteBstWouldAddEndPuncttrue
\mciteSetBstMidEndSepPunct{\mcitedefaultmidpunct}
{\mcitedefaultendpunct}{\mcitedefaultseppunct}\relax
\EndOfBibitem
\bibitem[Pan \latin{et~al.}(1989)Pan, Prokhorov, Komashko, Kaminsky,
  Kousenetsov, and Tretiatchenko]{92701}
Pan,~V.; Prokhorov,~V.; Komashko,~V.; Kaminsky,~G.; Kousenetsov,~M.;
  Tretiatchenko,~C. Superconducting properties of TaN and VN films. \emph{IEEE
  Transactions on Magnetics} \textbf{1989}, \emph{25}, 2000--2003\relax
\mciteBstWouldAddEndPuncttrue
\mciteSetBstMidEndSepPunct{\mcitedefaultmidpunct}
{\mcitedefaultendpunct}{\mcitedefaultseppunct}\relax
\EndOfBibitem
\bibitem[Romanov \latin{et~al.}(2018)Romanov, Zolotov, Vakhtomin, Divochiy, and
  Smirnov]{Romanov_2018}
Romanov,~N.~R.; Zolotov,~P.~I.; Vakhtomin,~Y.~B.; Divochiy,~A.~V.;
  Smirnov,~K.~V. Electron diffusivity measurements of VN superconducting
  single-photon detectors. \emph{Journal of Physics: Conference Series}
  \textbf{2018}, \emph{1124}, 051032\relax
\mciteBstWouldAddEndPuncttrue
\mciteSetBstMidEndSepPunct{\mcitedefaultmidpunct}
{\mcitedefaultendpunct}{\mcitedefaultseppunct}\relax
\EndOfBibitem
\bibitem[Saveskul \latin{et~al.}(2019)Saveskul, Titova, Baeva, Semenov,
  Lubenchenko, Saha, Reddy, Bogdanov, Marinero, Shalaev, Boltasseva, Khrapai,
  Kardakova, and Goltsman]{PhysRevApplied.12.054001}
Saveskul,~N.; Titova,~N.; Baeva,~E.; Semenov,~A.; Lubenchenko,~A.; Saha,~S.;
  Reddy,~H.; Bogdanov,~S.; Marinero,~E.; Shalaev,~V.; Boltasseva,~A.;
  Khrapai,~V.; Kardakova,~A.; Goltsman,~G. Superconductivity behavior in
  epitaxial $\mathrm{Ti}\mathrm{N}$ films points to surface magnetic disorder.
  \emph{Phys. Rev. Appl.} \textbf{2019}, \emph{12}, 054001\relax
\mciteBstWouldAddEndPuncttrue
\mciteSetBstMidEndSepPunct{\mcitedefaultmidpunct}
{\mcitedefaultendpunct}{\mcitedefaultseppunct}\relax
\EndOfBibitem
\bibitem[Jaim \latin{et~al.}(2015)Jaim, Aguilar, Sarabi, Rosen, Ramanayaka,
  Lock, Richardson, and Osborn]{6942188}
Jaim,~H. M.~I.; Aguilar,~J.~A.; Sarabi,~B.; Rosen,~Y.~J.; Ramanayaka,~A.~N.;
  Lock,~E.~H.; Richardson,~C. J.~K.; Osborn,~K.~D. Superconducting TiN films
  sputtered over a large range of substrate DC bias. \emph{IEEE Transactions on
  Applied Superconductivity} \textbf{2015}, \emph{25}, 1--5\relax
\mciteBstWouldAddEndPuncttrue
\mciteSetBstMidEndSepPunct{\mcitedefaultmidpunct}
{\mcitedefaultendpunct}{\mcitedefaultseppunct}\relax
\EndOfBibitem
\bibitem[vstanev1(2021)]{supercon}
vstanev1, Supercon. 2021; \url{https://github.com/vstanev1/Supercon}\relax
\mciteBstWouldAddEndPuncttrue
\mciteSetBstMidEndSepPunct{\mcitedefaultmidpunct}
{\mcitedefaultendpunct}{\mcitedefaultseppunct}\relax
\EndOfBibitem
\bibitem[Hamidieh(2018)]{hamidieh2018datadriven}
Hamidieh,~K. A data-driven statistical model for predicting the critical
  temperature of a superconductor. 2018\relax
\mciteBstWouldAddEndPuncttrue
\mciteSetBstMidEndSepPunct{\mcitedefaultmidpunct}
{\mcitedefaultendpunct}{\mcitedefaultseppunct}\relax
\EndOfBibitem
\end{mcitethebibliography}
% Produces the bibliography via BibTeX.

\end{document}